\documentclass[aps,amssymb,noshowpacs,superscriptaddress,nofootinbib]{revtex4}
\usepackage{graphicx}

\begin{document}
\title{Some Aspects of Holst and Nieh-Yan Terms in General Relativity with Torsion}
\author{Kinjal Banerjee}
\email{kinjal@iucaa.ernet.in}
\affiliation{IUCAA, Post Bag 4, Ganeshkhind, Pune - 411 007, India\\}
\date{\today}
\begin{abstract}

We explore the relation of the Holst term with the Nieh-Yan term in terms of metric variables. We show that the Holst term indeed affects the
classical equations of motion in the presence of matter with spin. Therefore the correct term to add to the Einstein-Hilbert action such that the
equations of motion are not affected is the Nieh-Yan term. We then calculate the torsion charge due to this term in the context of a perfect fluid
sphere with torsion and show that it vanishes once a horizon is formed but not otherwise. We also show that on adding torsion to General Relativity the 
Einstein's equations are no longer holographic in torsion although they continue to be so for the metric.
\end{abstract}

\maketitle

\section{Introduction}

In classical mechanics it is known that adding a total derivative to the action does not affect the classical equations of motion. The action from
which General Relativity can be obtained is the Einstein-Hilbert action. In 4 dimensions there exist 3 terms which can be be added to this action
such that the equations of motion are unaffected, the Gauss-Bonnet term \cite{gb}, the Pontryagin term \cite{cp} and the Holst term
\cite{holst},\cite{sa}. While the first two can be related to topological charges of the manifold, it was not clear till recently why the Holst term
does not modify the equations of motion. The Holst term however plays an important role in General Relativity because it has been showed that the
Einstein-Hilbert plus Holst action is the underlying action from which General relativity can be cast into a gauge theory form in terms of Ashtekar
variables \cite{ashtekar}.

It has been shown recently \cite{shyam},\cite{mercuri} that the Holst term is actually related to a total derivative known as the Nieh-Yan term 
\cite{niehyan} and has a topological interpretation. It has also been shown that, in the absence of matter, the canonical transformations generated by 
the Holst term and the Nieh-Yan term are the same. Hence the equations of motion are not affected by the addition of the Holst term even though it is not 
a total derivative. In recent works the Barbero-Immirzi parameter is considered as a field related to the 
$\theta$ parameter in QCD and the canonical formulation of fermionic theories have been constructed \cite{mercuri}. Most of these works have been done 
in the connection formulation. This is because the Holst term had been first introduced in the context of connection variables. Moreover based on
these variables there exists a background independent nonperturbative candidate theory of quantum gravity namely Loop Quantization Gravity(LQG) \cite{loop}. 

In this paper we clarify the relation between Holst and Nieh-Yan terms in terms of metric variables in the presence and absence of torsion. One
interesting feature of the Nieh-Yan term is that, in contrast to the Gauss-Bonnet and the  Pontryagin term, this is non trivial only in the presence
of torsion. A significant body of work exists in literature in General Relativity with torsion. Einstein-Sciama-Kibble theory is a theory of general 
relativity with torsion (see \cite{hehl} and references therein). Torsion also appears in the String Theory \cite{string} as the field strength of
the Kalb-Ramond field. In the light of the developments mentioned above, there may be further exploration on the role of torsion in theories which
may not necessarily be based on connection variables or quantized in the Loopy way. A clarification in terms of metric variables will help in further 
analysis of the known results in classical theories with torsion as well as facilitate a comparison of results coming from metric and connection theories, 
both classically and quantum mechanically . In that spirit, in this paper we show explicitly how the Holst term affects the equations of motion in the 
presence of matter and therefore should not be considered a canonical transformation. We also discuss a case where the total derivative term plays a role 
in the cosmological context. We also show that General Relativity no longer retains the holographic nature \cite{paddypr} when torsion is present. 

The paper is organized as follows.
In section II we recall the definitions of torsion and the curvature tensors in the presence of torsion. We then derive the relationship between
Holst and Nieh-Yan terms in metric variables. In particular we show that the Holst term is zero in the absence of torsion. This part of the paper is
of a pedagogical nature intended to explain the recent developments in metric variables and to clarify some of the misconceptions regarding the nature 
of the Holst term.
In section III we consider the equations of motion coming from the Einstein-Hilbert action along with the Holst term and show that the 
General Relativity result that torsion cannot exist without some matter distribution with spin may be violated with the addition of the Holst term.
In section IV we look at the Nieh-Yan term in the presence of a boundary. We calculate the topological charge due to torsion coming from this term
in the case of a perfect fluid with torsion with the outside spacetime being Schwarzschild. We show that the total derivative contributes a non zero 
torsion charge as long as the shell does not from a horizon.
In section V we show that the Einstein-Hilbert action does not retain its holographic structure \cite{paddyayan} in the presence of torsion even in
the case of ordinary General relativity and the addition of the Nieh-Yan term does not help in restoring holography. 
We conclude in section VI with a brief summary and future directions of work.

\section{Torsion and the Holst Term}

\subsection{Torsion}

In this section we shall first briefly recall how the concept of torsion is incorporated in General Relativity. We shall take the spacetime manifold 
to be ${\bf U}_4$ \cite{hehl}, i.e. a 4 dimensional metric spacetime with non zero torsion. 
Lowercase Latin letters ($a,b,\dots$) will refer to spacetime indices while uppercase Latin letters ($I,J,\dots$) will refer to internal indices.

Let $\nabla$ be the derivative operator compatible with the metric i.e. $\nabla_a g_{ab} = 0$. Then for any function $f$
\begin{eqnarray}
\bigg(\nabla_a \nabla_b  - \nabla_b \nabla_a\bigg)f &=& - \bigg(\Gamma_{~ab}^c - \Gamma_{~ba}^c \bigg) \nabla_c f \nonumber\\
&\equiv& - T_{~ab}^c \nabla_c f \label{metrictorsiondef}
\end{eqnarray}
where in the first line we have used $(\partial_a \partial_b = \partial_b \partial_a)$. The quantity $ T_{~ab}^c $ is known as the Torsion
tensor. It can be set to zero by demanding
\begin{eqnarray}
&&\bigg(\nabla_a \nabla_b  - \nabla_b \nabla_a\bigg)f = 0 \label{metrictorfree}\\
&& \mbox {i.e.} ~ ~ ~\Gamma_{~ab}^c = \Gamma_{~ba}^c  \nonumber
\end{eqnarray}
Equation (\ref{metrictorsiondef}) can be used to calculate the commutator of two vector fields ${\bf v}$ and ${\bf w}$. 
\begin{eqnarray}
\left[ {\bf v}, {\bf w} \right] (f) = \bigg( v^a \left( \nabla_a w^b \right) - w^a \left( \nabla_a v^b \right) \bigg) 
- v^a w^b  T_{~ab}^c \nabla_c f \label{veccom}
\end{eqnarray}
In particular, if we choose the vector fields to be basis vectors in a coordinate basis i.e. ${\bf v} = \partial/\partial x^a :={\bf x}$ and 
${\bf w} = \partial/\partial x^b :={\bf y}$  then from eqn. (\ref{veccom}) it is clear that
\begin{eqnarray}
\left[ {\bf x}, {\bf y} \right] (f) = 0 \label{veccomcod}
\end{eqnarray}
even in the presence of torsion because partial derivatives always commute. 

As an aside we can compare this with the notion of torsion using tetrads. Using the above equations we can find the commutation relations between 
a tetrad fields ($e^a_I$) which can be thought of as basis vector fields. They are  usually taken to form an orthogonal basis.
\begin{eqnarray}
e^a_I~ e_{a J} = \eta_{IJ} \nonumber 
\end{eqnarray}
Note that a coordinate basis is an orthogonal basis only for flat spacetimes in Cartesian coordinates.

Using eqn. (\ref{veccom}) we get
\begin{eqnarray}
(e_K)_a\left[e_I,e_J\right]^a = (e_K)_a \bigg( (e_I)^b \nabla_b (e_J)^a - (e_J)^b \nabla_b (e_I)^a\bigg) - (e_K)_c (e_I)^a (e_J)^b  T_{~ab}^c 
\end{eqnarray}
Defining connection 1-forms $\omega_{a IJ} = (e_I)^b \nabla_a (e_J)_b \Rightarrow \omega_{KIJ} =  (e_K)^a  (e_I)^b\nabla_a (e_J)_b$ we can rewrite
the above equation as
\begin{eqnarray}
(e_K)_a\left[e_I,e_J\right]^a = \bigg( \omega_{KIJ} -  \omega_{KJI} \bigg) - T_{KIJ} \label{tetradtor}
\end{eqnarray}
So torsion free condition is obtained by demanding
\begin{eqnarray}
\left[e_I,e_J\right]^a = \bigg( \omega_{KIJ} -  \omega_{KJI} \bigg) (e^K)^a \label{tetradtorfree}
\end{eqnarray}
In a coordinate basis the right hand side of the above equations are zero irrespective of the presence of torsion.

\subsection{Curvature Tensors in presence of Torsion}

It can be shown \cite{fabbri} that if we demand that all connections are metric, then without loss of generality we can assume that the torsion tensor 
is antisymmetric in all three indices. This is the case we shall mainly be interested in. Assuming that we can obtain the expressions of the curvature 
tensors which are modified from the standard expressions owing to the presence of torsion.

Writing the Riemann Christoffel symbols as 
\begin{eqnarray}
\Gamma_{~ab}^c = \hat\Gamma_{~ab}^c  + \gamma_{~ab}^c \label{rechs}
\end{eqnarray}
where $\hat\Gamma_{~ab}^c$ is symmetric and $ \gamma_{~ab}^c$ is antisymmetric in the lower two indices. Then from eqn 
(\ref{metrictorsiondef}) we have
\begin{eqnarray}
T_{~ab}^c ~ &=& ~2 \gamma_{~ab}^c \label{gammatorsion} 
\end{eqnarray}
Further, assuming totally antisymmetric torsion, the Riemann tensor can then be written as \cite{Eisenhart,fabbri}:
\begin{eqnarray}
 R_{a b c d} = \hat R_{a b c d} - \frac{1}{2}\bigg(\nabla_{c} T_{ a b d} - \nabla_{d} T_{ a b c }\bigg) + 
 \frac{1}{4}\bigg( T_{a e c} T^{e}_{~b d} - T_{a e d} T^{e}_{~ b c} \bigg)  \label{riemann}
\end{eqnarray}
where the hatted quantities are the corresponding tensors in the absence of torsion. The Ricci tensor is given by:
\begin{eqnarray}
	R_{a b} = \hat R_{a b} +\frac{1}{2} \nabla_e T^{e}_{~ a b} - \frac{1}{4} T^{e}_{~ f b} T^{f}_{~ b e} \label{ricciT}
\end{eqnarray}
while the Ricci scalar is given by
\begin{eqnarray}
R = \hat R -\frac{1}{4} T_{a e d} T^{a e d} \label{ricci}
\end{eqnarray}

{\bf Note}
In literature \cite{hehl,fabbri}, torsion is usually incorporated through the {\em contorsion} tensor which is given by
\begin{eqnarray}
K^a_{~ b c} := T^a_{~ b c} + T^{~ ~a}_{b c} + T^{~ ~a}_{ c b} \label{contorsion}
\end{eqnarray}
This reduces to the torsion tensor when torsion is antisymmetric in the lower two indices. Since in this paper we will be dealing with totally
antisymmetric (or at least antisymmetric in the two lower indices) torsion, we will express our results in terms of the torsion tensor.

\subsection{Holst Term}

Having obtained the curvature scalars in the presence of torsion we can begin analysis of the Holst term. In terms of the tetrads and the curvature 
tensors the Holst term can be written as \cite{holst}
\begin{eqnarray}
S_H &=& \alpha \int \mbox{d}^4 x ~e e^a_I e^b_J \epsilon^{IJKL} R_{a b KL}  \label{holst}
\end{eqnarray}
where $e=\sqrt{-g}~$ and $\alpha$ is related to the Barbero-Immirzi parameter \cite{shyam}. Since our objective is to work in terms of metric
variables we can rewrite eqn. (\ref{holst}) as
\begin{eqnarray}
&=& \alpha \int \mbox{d}^4 x ~e e^{a}_I e^{b}_J e^{c}_K e^{d}_L \epsilon^{IJKL} R_{abcd}\nonumber\\
&=& \alpha \int \mbox{d}^4 x ~ \epsilon^{abcd} ~ R_{abcd}  \label{metricholst}
\end{eqnarray}
where $ \epsilon^{abcd} = \sqrt{-g}~ \eta^{abcd}$ with $\eta^{abcd}$ being the Levi-Civita symbol.

Note that in the absence of torsion this term vanishes from cyclic identity of the Riemann tensor. So unlike the Euler and the Pontryagin terms this
term is non trivial only when torsion is present. In the presence of torsion  we have to use eqn (\ref{riemann}) as the definition of Riemann tensor in 
eqn (\ref{metricholst}).

The first term coming from eqn (\ref{riemann}) is the Riemann tensor without torsion which again becomes zero from cyclic identity. The other two terms are 
non zero. In particular the second group of terms give:
\begin{eqnarray}
-\partial_a\bigg(\epsilon^{a b c d} T_{b c d} \bigg)
\end{eqnarray}
while the third group gives
\begin{eqnarray}
\epsilon^{a b c d} T^{e} _{~ a c} T_{e b d} 
\end{eqnarray}
Therefore using eqn (\ref{riemann}) as the definition for $R_{abcd}$ and contracting it with $\epsilon^{abcd}$ we arrive at the following equation:
\begin{eqnarray}
\partial_a\bigg(\epsilon^{a b c d} T_{b c d} \bigg) =  \frac{1}{2} \epsilon^{a b c d} T^{e} _{~ a c} T_{e b d} - \epsilon^{a b c d} R_{a b c d}
\label{nyh}
\end{eqnarray}
The term on the R.H.S. is known as the Nieh-Yan term and is equal to a total derivative. We can see that the Holst term we started with is related to a
total derivative along with a term quadratic in torsion. In the absence of torsion, the quadratic torsion term becomes 
zero and the addition of the Holst term is equivalent to the addition of a total derivative. Therefore the equations of motion are not affected even
when we add to the action, a term which is not a total derivative. In \cite{shyam} torsion was set to zero by strongly imposing a pair of second class
constraints. Because it is not a total derivative the naive addition of the Holst term will change the classical equations of motion in the presence
of torsion as we will see in the next section.


\section{Holst Term and Equations of Motion}

In this section we will look at the equations of motion coming from an action which has the Holst term added to the Einstein Hilbert action. The metric 
$g_{ab}$ and torsion $T^a_{~ b c}$ are the two independent basic variables. We take the gravitation part of our action to be
\begin{eqnarray}
S_{grav} &=& S_{EH} + 2 \alpha S_{H} \nonumber \\
&=& \int \mbox{d}^4 ~x ~ \sqrt{-g} R  ~ ~ + ~ ~  2 \alpha \int \mbox{d}^4 x ~ \epsilon^{abcd} ~ R_{abcd}  \label{fullaction1}
\end{eqnarray}
The variation of the Holst term is then equivalent to the variation of $\frac{1}{2}\epsilon^{a b c d} T^{e} _{~ a c} T_{e b d}$ because the other term 
is a total derivative. Therefore we can take our action to be
\begin{eqnarray}
S_{grav} = \int \mbox{d}^4 ~x ~ \sqrt{-g} R  ~ ~ + ~ ~  
\alpha\int \mbox{d}^4 x ~\epsilon^{a b c d} T^{e} _{~ a c} T_{e b d}\label{fullaction2}
\end{eqnarray}
Note that the second term is not trivially zero from symmetry properties. 

The variation of $T_{~ab}^c$  using eqn (\ref{gammatorsion}):
\begin{eqnarray}
T_{~ab}^c ~ = ~2 \gamma_{~ab}^c ~ ~ ~
\Rightarrow ~ ~ ~ (\delta T_{~ab}^c) = 2 (\delta \gamma_{~ab}^c)
\end{eqnarray}
And 
\begin{eqnarray}
T_{cab} = g_{cd} T_{~ab}^d ~ ~ ~
\Rightarrow ~ ~ ~  (\delta T_{cab} ) = 2 (\delta  g_{cd}) \gamma_{~ab}^d + 2 g_{cd} (\delta \gamma_{~ab}^d)
\end{eqnarray}
Using these we can write
\begin{eqnarray}
(\delta S_H) &=& \eta^{a b c d} \left(\delta ~ \sqrt{-g} ~ T^{e} _{a c} T_{e b d}\right) \nonumber\\
&=& 4 \eta^{a b c d} \sqrt{-g}~\bigg[ -\frac{3}{2} \gamma_{ac}^{~ ~e} \gamma_{ebd} ~ g_{ij} ~(\delta g^{ij}) 
+ 2  \gamma_{ebd} (\delta  \gamma_{ac}^{~~e}) \bigg] \nonumber\\
&=& 2\eta^{a b c d} \sqrt{-g}~\bigg[ -\frac{3}{4} T_{ac}^{~ ~e} T_{ebd} ~ g_{ij} ~(\delta g^{ij}) 
+ T_{ebd} (\delta  T_{ac}^{~~e}) \bigg]
\end{eqnarray}

The variation of the Einstein-Hilbert is well known \cite{hehl}
\begin{eqnarray}
(\delta S_{EH}) &=& \sqrt{-g}~ \bigg( \hat R_{a b} - \frac{1}{2}\hat R g_{ab}\bigg)  (\delta g^{ab})  \nonumber\\
&+&  \sqrt{-g}\Bigg\{ 2 \gamma_a^{~ c e}\gamma_{ebc} (\delta g^{ab}) 
- \frac{3}{2} \gamma_{a c}^{~ ~ e}\gamma_e^{~ a c} ~ g_{ij} (\delta g^{ij}) + 2 \gamma_e^{~ a c} (\delta \gamma_{ac}^{~~e}) \bigg\} \nonumber \\
 &=& \sqrt{-g}~ \bigg( \hat R_{a b} - \frac{1}{2}\hat R g_{ab}\bigg)  (\delta g^{ab})  \nonumber\\
&+&  \frac{\sqrt{-g}}{2}\Bigg\{ T_a^{~ c e} T_{ebc} (\delta g^{ab}) 
- \frac{3}{4} T_{a c}^{~ ~ e} T_e^{~ a c} ~ g_{ij} (\delta g^{ij}) + T_e^{~ a c} (\delta T_{ac}^{~~e}) \bigg\}
\end{eqnarray}
Therefore the variation of the full action can be expressed as
\begin{eqnarray}
(\delta S_{grav}) &=& (\delta S_{EH}) + 2\alpha(\delta S_{H}) \nonumber \\
&=& \sqrt{-g}~\bigg(\hat R_{a b} - \frac{1}{2}\hat R g_{ab}\bigg)  (\delta g^{ab}) 
+ \frac{\sqrt{-g}}{2} \Bigg( T_a^{~ c e} T_{ebc}(\delta g^{ab}) \nonumber\\ 
&&  -\frac{3}{4}\bigg[ T_{a c}^{~ ~ e} T_e^{~ a c}  +4\alpha \eta^{a b c d} T_{ac}^{~ ~e} T_{ebd} \bigg] g_{ij}  (\delta g^{ij})
+ \bigg[T_e^{~ a c} +  4 \alpha \eta^{a b c d} T_{ebd}\bigg] (\delta T_{ac}^{~~e}) \Bigg) \label{actionvariation}
\end{eqnarray}

We are interested in the case when matter is present. In the presence of a matter distribution with the total action given by 
\begin{eqnarray}
S =  \int \mbox{d}^4 ~x ~ \left( \frac{1}{\kappa} S_{grav} + S_{matter} \right)
\end{eqnarray}
where $\kappa = 16 \pi G$. The {\em energy-momentum} tensor and a {\em spin energy} tensor is usually defined by
\begin{eqnarray}
T_{ab} &=& \frac{1}{ \sqrt{-g}} \frac{\delta S_{matter}}{\delta g_{ab}} \label{emdef} \\
S_{a}^{~ b c} &=& \frac{1}{ \sqrt{-g}} \frac{\delta S_{matter}}{\delta T^a_{~ b c}} \label{spindef}
\end{eqnarray}
Using these definitions and eqn (\ref{actionvariation}) we find
\begin{eqnarray}
\kappa T_{ab} &=& \bigg(\hat R_{a b} - \frac{1}{2}\hat R g_{ab}\bigg) + \frac{1}{2} T_a^{~ c e} T_{ebc} 
-\frac{3}{8}\bigg( T_{f c}^{~ ~ e} T_e^{~ f c}  + 4\alpha \eta^{f b c d} T_{fc}^{~ ~e} T_{ebd} \bigg) g_{ab} \label{tab} \\
\kappa S_{e}^{~ a c} &=& \frac{1}{2} \bigg(T_e^{~ a c} +  4 \alpha \eta^{a b c d} T_{ebd}\bigg) \label{sabc}
\end{eqnarray}
If the Holst term is not there $\alpha=0$ a couple of points can be noted from the above equations:\\
a) In the absence of spin $S_{e}^{~ a c}$ in the matter distribution $T_{e}^{~ a c}=0$ from eqn (\ref{sabc}). Thus there can be no torsion
without spin and we get back standard Einstein's equations with the standard definition for $T_{ab}$. This also ensures that torsion cannot propagate
from a region of non zero spin distribution to a vacuum region.\\
b) If spin is not zero we can solve  eqn (\ref{sabc}) to obtain $T_{e}^{~ a c} = 2\kappa S_e^{~ a c}$ which can then be substituted
into eqn (\ref{tab}) to obtain a modified energy momentum tensor $\tilde T_{ab}$
\begin{eqnarray}
\kappa \tilde T_{ab} &=& \bigg(\hat R_{a b} - \frac{1}{2}\hat R g_{ab}\bigg) \nonumber \\
&=& T_{ab} + \kappa\bigg( 2 S_a^{~ c e} S_{b e c} +\frac{3}{2} S_{fce} S^{fce} \bigg) 
\end{eqnarray}

However in the presence of the Holst term, solving eqn (\ref{sabc}) is no longer trivial. After some algebra it can be shown that
\begin{eqnarray}
T_{e a c} = \frac{1}{4(1+64 \alpha^2)} \bigg(S_{e a c} -2 \alpha S_e^{~ b d} \eta_{b a d c}\bigg)
\end{eqnarray}
This is the expression which now has to be substituted into  eqn (\ref{tab}) to obtain a modified energy momentum tensor. The expression of
the modified energy momentum tensor will be different in the presence of the Holst term. This also shows why the Barbero-Immirzi parameter shows up in 
the classical equations of motion when matter is included \cite{rovelli}. In fact it is no longer clear that torsion cannot exist
without spin. The modified energy momentum tensor is different in the presence of the Holst term and depending on the matter action, the
Holst term needs to be modified if we want to ensure that the modified energy momentum tensor remains unchanged. This modification has to be done on
a case by case basis and is therefore not universal.

These show that if we consider theories of gravitation with matter, where the independent variables in the gravitational action are the metric and
torsion, a theory based on $S_{EH}$ and one based on $S_{EH} + S_{H}$ are not equivalent even classically. Of course, since the Holst term is not a
total derivative there is no reason for this to happen. The term we can add to the action without affecting the classical equations of motion is
actually the Nieh-Yan term.

\section{Nieh-Yan Term and Topological charge}

We know that the Gauss-Bonnet term and the Pontryagin term can be added to the Einstein Hilbert action in four dimensions without affecting the
equations of motion. On integration these give a number characteristic to the manifold. The Gauss-Bonnet term gives the Euler characteristic while
the Pontryagin term gives the Pontryagin class of the manifold over which the integral is taken. In this section we shall try to see what the
Nieh-Yan term gives when we consider a physical situation in a manifold with a boundary.

An obvious example of a spacetime with boundary is a Black Hole. One widely studied case of the formation of a Black Hole in the presence of torsion 
is the case of a collapsing Weyssenhoff fluid sphere \cite{torsionfluid}. 

Weyssenhoff fluid is a perfect fluid with spin whose spin tensor is given by
\begin{eqnarray}
S^i_{~ j k} &=& u^i s_{jk} \\
\mbox{with} ~ ~ u^k s_{jk} &=& 0
\end{eqnarray}
where $u^i$ is the 4-velocity of the fluid and $s_{jk}$ is the intrinsic angular momentum. 

The energy momentum tensor of the fluid respecting the conservation laws of spin and energy momentum is given by
\begin{eqnarray}
T_{jk} &=& h_j u_k - P g_{jk} \\
\mbox{where} ~ ~ h_j &=& (\rho + P) u_j - u^i \nabla_k ( u^k s_{ij} ) 
\end{eqnarray}
where $P$ and $\rho$ are the pressure and density of the fluid. If the metric is spherically symmetric and the intrinsic angular momentum is aligned
in the $r$ direction the only non zero components of $s_{ij}$ are
\begin{eqnarray}
s_{ij} ~=~K~=~ -s_{ji} \nonumber
\end{eqnarray}
If the fluid is assumed to be static $u^i = \delta^i_0$. Then the only non zero components of the spin tensor are
\begin{eqnarray}
S^0_{~ 2 3} ~ = ~ K ~ = ~ - S^0_{~ 3 2 }
\end{eqnarray}
It has been shown \cite{prasanna} that it is possible to construct a spherically symmetric Weyssenhoff fluid distribution with the outside vacuum 
spacetime metric given by the Schwarzschild metric consistent with the boundary conditions \cite{boundary}. The boundary conditions are basically
equivalent to the statement that the {\em effective pressure} $P'$ on the surface is zero, where the effective pressure is given by
\begin{eqnarray}
P' = P - \frac{\kappa}{8} K^2
\end{eqnarray}
The physical situation corresponds to a fluid sphere of spinning matter while the spacetime outside the sphere is vacuum and without
torsion. Note that the fluid is assumed to have intrinsic spin which is different from the case a rotating fluid sphere. Since the outside spacetime
is vacuum from eqn (\ref{sabc}) (with $\alpha=0$) we can see that there is no torsion.

Let the fluid sphere be of radius $r=a$. Then from Birkhoff's theorem, for $r>a$ the metric is the standard Schwarzschild metric
\begin{eqnarray}
\mbox{d} s^2 = -\left(1 -\frac{2M}{r}\right)\mbox{d}t^2 + \left(1 -\frac{2M}{r}\right)^{-1}\mbox{d}r^2 +r^2 \mbox{d} \theta^2 
+r^2 \sin^2 \theta \mbox{d} \phi^2  \label{outmetric}
\end{eqnarray}
while the metric inside the sphere is given by
\begin{eqnarray}
\mbox{d} s^2 = &&-\left(\frac{3}{2}\left[1-\frac{2M}{a}\right]^{\frac{1}{2}} - \frac{1}{2}\left[1 -\frac{2M r^2}{a^3}\right]^{\frac{1}{2}}\right)
\mbox{d}t^2 \label{inmetric}\\ 
&& ~ ~ + \left(1 -\frac{2M}{r}\right)^{-1}\mbox{d}r^2 + r^2 \mbox{d} \theta^2 + r^2 \sin^2 \theta \mbox{d} \phi^2  \nonumber
\end{eqnarray}
The inside and the outside metrics match on the surface $r=a$. However the fluid sphere is not of uniform density. Also from the metric inside we can
see that the singularity at $r=0$ can only occur if $M/a = \frac{4}{9}$.

From the equations of motion (\ref{sabc}) which are unaffected by the addition of a total derivative, we can determine the nonzero 
torsion components inside the sphere 
\begin{eqnarray}
T^0_{~ 2 3} ~ = ~2 \kappa K ~ = ~ - T^0_{~ 3 2 }
\end{eqnarray}
Let us consider the Nieh-Yan action inside the shell: 
\begin{eqnarray}
S_{NY} = \alpha \int_{\mbox{in}} \mbox{d}^4 x ~ \partial_a\bigg(\epsilon^{a b c d} T_{b c d} \bigg) \label{ny}
\end{eqnarray}
Evaluating this  on the surface of the fluid sphere, $r=a$ hypersurface:
\begin{eqnarray}
S_{NY} &=& - 2 \alpha \int \mbox{d} t ~ \mbox{d} r ~ \mbox{d} \theta ~\mbox{d} \phi \partial_r\bigg(\sqrt{-g} ~ g_{00} T^0_{~ 2 3}\bigg) \nonumber \\
&=& - 4 \kappa \alpha \int_{r=a} \mbox{d} t ~ \mbox{d} \theta ~\mbox{d} \phi \bigg(\sqrt{-g}~ g_{00} K \bigg) \nonumber \\
&=& - 4 \kappa \alpha \int_{r=a} \mbox{d} t ~ \mbox{d} \theta ~\mbox{d} \phi r^2 \sin^2 \theta \left(1 -\frac{2M}{r}\right) K
\end{eqnarray}

This can be easily evaluated by looking at the Euclidean case. Note that the topology of the Euclidean Black Hole is 
${\mathbb R}^2 \times S^2$ while the topology of the boundary is $S^1 \times S^2$. The proper length along $S^1$ is given by $\beta$ the inverse of the
temperature at the boundary. Using the technique of Euclidean continuation without the presence of a horizon from \cite{frolov} we can calculate that 
$\beta$ at the boundary $r=a$ is given by
\begin{eqnarray}
\beta = 8 \pi M \left( 1 - \frac{2M}{a} \right)^{\frac{1}{2}}
\end{eqnarray}
Therefore, at $r=a$
\begin{eqnarray}
S_{NY} = -128 \pi^2 \alpha \kappa M a^2 K \left( 1 - \frac{2M}{a} \right)^{\frac{3}{2}}
\end{eqnarray}
Clearly this vanishes only when $a=2M$ i.e. when the collapsing fluid forms a horizon but is non-zero when the horizon is not yet formed. The surface
term therefore contributes a {\em torsion charge} in a manifold with a boundary, in this case the surface of the sphere. However it does not contribute to
the entropy of the Black Hole. Note that we are interested in the region outside the sphere with no torsion with the boundary given by the surface of
the sphere. The equations of motion in the interior of the sphere are used to determine the value of torsion on the boundary. 

The reason why the contribution of the total derivative term does not remain constant in this case is because the basic variable in our equations of
motion $T^a_{~bc}$ while in the Nieh-Yan term torsion occurs as $T_{abc}$ which involves the metric. One interesting case where we do get a constant is 
if there is a singularity at $r=0$, the ratio between $M$ and $a$ is a constant number and we get a constant contribution from the total derivative on 
the boundary. 

Interestingly the contribution of the Nieh-Yan term is thus non zero when there is a boundary but vanishes when the boundary is a causal boundary i.e. 
a horizon. This leads on to the question whether the Einstein's equation retain their holographic property in the presence of torsion.


\section{Torsion and Holography}

To discuss the question posed at the end of the last section let us briefly recall the concept of holography in gravitational action.
One interesting feature of the Einstein Hilbert action is that can be split into a bulk part and a total derivative \cite{paddyayan}. That is
\begin{eqnarray}
&& \sqrt{-g} R = \sqrt{-g} L_{quad} -\partial_c P^c ~~~~~~~~ \mbox{where} \label{ehlqpc} \\
&& L_{quad} = g^{ab}\bigg(\Gamma^i_{~ja} \Gamma^j_{~ib} - \Gamma^i_{~ab} \Gamma^j_{~ij}\bigg) \label{lquad} \\
&& P^c =  \sqrt{-g} \bigg( g^{ck}  \Gamma^m_{~km} - g^{ik} \Gamma^c_{~ik} \bigg) \label{pc}
\end{eqnarray}
The equations of motion of $L_{quad}$ are equivalent to the equations of motion of the $L_{EH}$. In other words, if we start with a theory based 
on $L_{quad}$ we will end with the same equations of motion as coming from General Relativity. The two terms are related by what is known as
{\em holography} \cite{paddypr}:
\begin{eqnarray}
\sqrt{-g} \partial_c P^c = - \partial_a \left( g_{ij} 
 \frac{\partial \sqrt{-g} L_{quad}}{\partial \left(\partial _a g_{ij} \right)}\right) \label{holo}
\end{eqnarray}
These results have been obtained when there is no torsion. In the presence of torsion recall the definition of Riemann-Christoffel symbols from 
eqn (\ref{rechs}).
\begin{eqnarray}
\Gamma_{~ab}^c = \hat\Gamma_{~ab}^c  + \frac{1}{2} T_{~ab}^c \nonumber 
\end{eqnarray}
Let us split the various terms in eqn (\ref{holo}) as 
\begin{eqnarray}
L_{quad} &=& \hat L_{quad} + \tilde L_{quad} \\
P^c &=& \hat P^c + \tilde P^c 
\end{eqnarray}
where the hatted quantities come from the symmetric part of the connection and the tilde terms depend on torsion. Then
\begin{eqnarray}
\tilde L_{quad} &=& \frac{1}{4} g^{ab} \bigg(T^i_{~ja} T^j_{~ib} - T^i_{~ab} T^j_{~ij}\bigg) \nonumber\\
&=& \frac{1}{4} g^{ab} T^i_{~ja} T^j_{~ib}  \nonumber\\
&=& - \frac{1}{4} T^{ijb} T_{ijb} \label{lquadtor}
\end{eqnarray}
Similarly
\begin{eqnarray}
\tilde P^c =  \sqrt{-g} \bigg( g^{ck} T^m_{~km} - g^{ik} T^c_{~ik} \bigg) ~=~ 0 \label{pctor}
\end{eqnarray}
In both the cases we have used the fact that we are interested in totally antisymmetric torsion. Then 
\begin{eqnarray}
&& \bigg(\sqrt{-g} \hat L_{quad} -\partial_c \hat P^c\bigg) +  \bigg(\sqrt{-g} \tilde L_{quad} -\partial_c \tilde P^c \bigg) \nonumber \\
&=&  \sqrt{-g} \left( \hat R -  \frac{1}{4} T^{ijb} T_{ijb} \right) \\
&=&  \sqrt{-g} R  ~~~~~~~~~~~~~ \mbox{   from eqn (\ref{ricci})}
\end{eqnarray}
Thus, the decomposition of $L_{EH}$ in the form given in eqn (\ref{ehlqpc}) goes through in the presence of torsion. Note that the extra piece due to
torsion in the Ricci scalar comes from $L_{quad}$ while the surface term has no extra contribution. Since there are no derivatives of torsion in the
action the corresponding holographic equation for torsion is trivial. In the holographic formulation, it is the surface term which leads to Black
Hole entropy and the inclusion of torsion does not change the surface term indicating that the entropy of a Black Hole is independent of the type of 
matter which has collapsed to form it. In the previous section we had seen that the inclusion of the Nieh-Yan term did not affect Black Hole entropy
(at least in the case we studied). 

Let us try to see how the Nieh-Yan surface term differs from a `holographic' surface term. The total derivative can be rewritten as
\begin{eqnarray}
L_{NY} = \partial_a\bigg(\epsilon^{a b c d} T_{b c d} \bigg) 
= \sqrt{-g} ~\eta ^{a b c d} \partial_a ( T_{b c d}) +  \eta ^{a b c d} T_{b c d}  \partial_a (\sqrt{-g})
\end{eqnarray}
Now consider the torsion dependent parts of this term and the Einstein-Hilbert Lagrangian
\begin{eqnarray}
L_{grav}^{tor} =  - \frac{\sqrt{-g}}{4} T^{abc} T_{abc} + \alpha  \sqrt{-g}~ \eta ^{a b c d} \partial_a ( T_{b c d}) + 
\alpha \eta ^{a b c d} T_{b c d}  \partial_a (\sqrt{-g})
\end{eqnarray}
Note that this Lagrangian also has a term quadratic in the field variables and a surface term. We can define
\begin{eqnarray}
\Pi^{abcd} = \frac{\partial L_{grav}^{tor}}{\partial(\partial_a ( T_{b c d}))} = \alpha  \sqrt{-g} \eta ^{a b c d}
\end{eqnarray}
which allows us to rewrite the Nieh-Yan term as
\begin{eqnarray}
\alpha L_{NY} = \partial_a \bigg(\Pi^{abcd}  T_{b c d} \bigg) =  \partial_a 
\left(\frac{\partial L_{grav}^{tor}}{\partial(\partial_a ( T_{b c d}))}   T_{b c d} \right)
\end{eqnarray}
Now the bulk term dependent on torsion is $ - \frac{\sqrt{-g}}{4} T^{abc} T_{abc}$. For this, a term similar to the holographic term for the metric 
degrees of freedom will look like
\begin{eqnarray}
\partial_a \left( \frac{T_{ijk}}{4} \frac{\partial \left(\sqrt{-g}T^{abc} T_{abc}\right) }{\partial \left(\partial _a T_{ijk} \right)}\right)
\end{eqnarray}
which is not similar to the Nieh-Yan term. The action is therefore not holographic for torsion even with the addition of the Nieh-Yan term. 
We see that although we started with metric and torsion as the two independent variables, the action is holographic in the sense of eqn (\ref{holo}) 
in metric but not in torsion.

\section{Discussion}

We have explored some features of the Nieh-Yan term from the perspective of metric variables in this paper. The first significant distinction of this
term is that, unlike the two other topological invariants in 4 dimensions, this becomes zero in the absence of torsion. Therefore the effect of the
Nieh-Yan term can only be studied on ${\bf U}_4$ manifolds. On a ${\bf U}_4$ manifold the Nieh-Yan term is related to the Holst term and a term
quadratic in torsion. It is shown that simply adding the Holst term then affects the classical equations of motion. Significantly it keeps the
possibility of torsion existing without spin open, unlike in General Relativity. It might be interesting to find whether any such solutions exist but
such solutions, if they exist will be solutions of an action which is not related to Einstein-Hilbert action by any canonical transformations.

We then studied one particular case where the Nieh-Yan term is non zero on a boundary of a spacetime. However the {\em topological charge} vanishes when
the boundary is a horizon. If there exists other examples of solutions of Einstein-Cartan theory with a boundary, it will be interesting to study
whether this effect is generic or a special feature of the this particular solution. Interestingly, the addition of Gauss-Bonnet invariant to the 
Einstein-Hilbert action was considered in the context of local AdS asymptotic geometry \cite{zanelli}. It was shown that although the surface term does 
not modify the bulk field equations, it changes the conserved currents and, therefore, the definition of conserved charges of the theory. In our
calculation we have used results which respect the conservation laws of matter and spin coming from General Relativity with torsion. Whether
something similar happens on the addition of Nieh-Yan term needs to be further investigated.

From the point of view of holography, it is clear that metric and torsion have to be considered on different footings even with the inclusion of an
explicit boundary term which depends only on torsion. It emphasizes that the holographic nature of Einstein's equations is a special and distinctive
feature of metric theories and does not hold on arbitrary extensions of the theory. However, notably the holography in metric continues to hold even
in the presence of torsion although the total action is not holographic.

In general, a topological term added to the action may show up in the quantum theory. For example in \cite{olea} it was shown that the addition of 
topological invariants of the Euler class is equivalent to holographic renormalization procedure in AdS/CFT. In some particular cases (eg. QCD with 
massive fermions) a topological term is characterizes a nonperturbative vacuum structure of the theory with a non zero tunnelling probability between 
various ground states. To see such an effect in quantum gravity, coming from the Nieh-Yan term  we will have to first study some the path integral 
formulation which incorporates torsion along with the metric. If such a formulation is possible it may be possible to explore whether the Barbero-Immirzi
parameter plays any role in labelling the ground states in a quantum theory of gravity.

{\bf Acknowledgements:}
I would like to thank T. Padmanabhan for suggesting the problem and for the discussions, his comments and his careful reading of the manuscript. 
I would also like to thank Ghanashyam Date for reading the manuscript and his queries and comments. 
Thanks are also due to Dawood Kothawala, Alok Laddha and Sandipan Sengupta for discussions.


\end{document}